\documentclass[nofootinbib,amsmath,twocolumn,notitlepage,preprintnumbers]{revtex4-1}
\usepackage{multirow}
\usepackage{amssymb,esvect,amsmath,graphicx,latexsym,amsthm,slashed,eso-pic}
\usepackage{hyperref}

\newcommand{\ie}{{\it i.e.}}

\newcommand{\beq}{\begin{equation}} \newcommand{\eeq}{\end{equation}}
\newcommand{\bea}{\begin{eqnarray}} \newcommand{\eea}{\end{eqnarray}}

\def\lsim{\mathrel{\raise.3ex\hbox{$<$\kern-.75em\lower1ex\hbox{$\sim$}}}}
\def\gsim{\mathrel{\raise.3ex\hbox{$>$\kern-.75em\lower1ex\hbox{$\sim$}}}}

\newcommand{\be}{\begin{eqnarray}}
\newcommand{\ee}{\end{eqnarray}}

\newcommand{\benum}{\begin{enumerate}}
\newcommand{\eenum}{\end{enumerate}}
\newcommand{\bi}{\begin{itemize}}
\newcommand{\ei}{\end{itemize}}

\newcommand{\neff}{{N_{\rm eff}}}

\begin{document}

\preprint{FERMILAB-PUB-18-066-A}

\title{Severely Constraining Dark Matter Interpretations of the 21-cm Anomaly}
\author{Asher Berlin$^{a}$}
\thanks{ORCID: http://orcid.org/0000-0002-1156-1482}
\author{Dan Hooper$^{b,c,d}$}
\thanks{ORCID: http://orcid.org/0000-0001-8837-4127}
\author{Gordan Krnjaic$^b$}
\thanks{ORCID: http://orcid.org/0000-0001-7420-9577}
\author{Samuel D.~McDermott$^b$}
\thanks{ORCID: http://orcid.org/0000-0001-5513-1938}

\affiliation{$^a$SLAC National Accelerator Laboratory, Menlo Park CA, 94025, USA}
\affiliation{$^b$Fermi National Accelerator Laboratory, Theoretical Astrophysics Group, Batavia, IL, USA}
\affiliation{$^c$University of Chicago, Kavli Institute for Cosmological Physics, Chicago IL, USA}
\affiliation{$^d$University of Chicago, Department of Astronomy and Astrophysics, Chicago IL, USA}

\date{\today}

\begin{abstract}

The EDGES Collaboration has recently reported the detection of a stronger-than-expected absorption feature in the global 21-cm spectrum, centered at a frequency corresponding to a redshift of $z\sim 17$. This observation has been interpreted as evidence that the gas was cooled during this era as a result of scattering with dark matter. In this study, we explore this possibility, applying constraints from the cosmic microwave background, light element abundances, Supernova 1987A, and a variety of laboratory experiments. After taking these constraints into account, we find that the vast majority of the parameter space capable of generating the observed 21-cm signal is ruled out. The only range of models that remains viable is that in which a small fraction, $\sim0.3-2\%$, of the dark matter consists of particles with a mass of $\sim10-80$ MeV and which couple to the photon through a small electric charge, $\epsilon \sim 10^{-6}-10^{-4}$. Furthermore, in order to avoid being overproduced in the early universe, such models must be supplemented with an additional depletion mechanism, such as annihilations through a $L_{\mu}-L_{\tau}$ gauge boson or annihilations to a pair of rapidly decaying hidden sector scalars.

\end{abstract}

\maketitle

\section{Introduction}

The Experiment to Detect the Global Epoch of reionization Signature (EDGES) Collaboration~\cite{nature} has recently reported the measurement of a feature in the absorption profile of the sky-averaged radio spectrum, centered at a frequency of 78 MHz and with an amplitude of 0.5 K. Although such a feature was anticipated to result from the 21-cm transition of atomic hydrogen (at $z\sim 17$), the measured amplitude of this signal is significantly larger than expected, at a confidence level of 3.8$\sigma$. If confirmed, this measurement would indicate that either the gas was much colder during the dark ages than expected, or that the temperature of the background radiation was much hotter. It has been argued~\cite{nature2} that standard astrophysical mechanisms~\cite{Cohen:2016jbh,nature,nature2,Fialkov:2018xre,Feng:2018rje,Ewall-Wice:2018bzf,Mirocha:2015jra} cannot account for this discrepancy, and that the only plausible explanations for this observation are those which rely on interactions between the primordial gas and light dark matter particles, resulting in a significant cooling of the gas~\cite{nature,nature2,Fialkov:2018xre} (see also Refs.~\cite{Tashiro:2014tsa,Munoz:2015bca}).

In order for the dark matter to cool the gas efficiently, it must have some rather specific characteristics. Firstly, equipartition requires that the dark matter particles be fairly light, with masses no larger than a few GeV. Secondly, if the cross section for dark matter scattering with gas is independent of velocity, a variety of constraints, including those from observations of the cosmic microwave background (CMB), would restrict the couplings to well below the values required to explain the observed amplitude of the absorption feature. Velocity-dependent scattering can relax such constraints, however, as the average velocities of baryons and dark matter particles were at approximately their minimum value during the cosmic dark ages (due to higher temperatures and having fallen into the gravitational potential of dark matter halos at earlier and later times, respectively). With this in mind, we can maximize the impact of dark matter-baryon scattering during this era by considering models in which $\sigma (v) \propto v^{-4}$. In terms of model building, this consideration directs us towards models in which the dark matter-baryon interactions are mediated by a particle that is much lighter than the temperature at $z \sim 17$, corresponding to $m_{\rm med} \lsim 10^{-3}$ eV. Experimental constraints on fifth forces are very stringent in this mass range~\cite{Adelberger:2006dh,Kapner:2006si}, however, and rule out a new light mediator with couplings in the range required to explain the observed absorption feature. Furthermore, such a particle would invariably contribute to the energy density of radiation during recombination at a level well above current constraints~\cite{Ade:2015xua}. In light of these considerations, the only options available are models in which the dark matter carries a small quantity of electric charge (\ie~a millicharge), and thus couples weakly to the photon~\cite{McDermott:2010pa,Davidson:2000hf,Chuzhoy:2008zy,Dvorkin:2013cea,Dolgov:2013una,Dubovsky:2003yn,Feldman:2007wj}.

\begin{figure*}
\includegraphics[width=3.2in,angle=0]{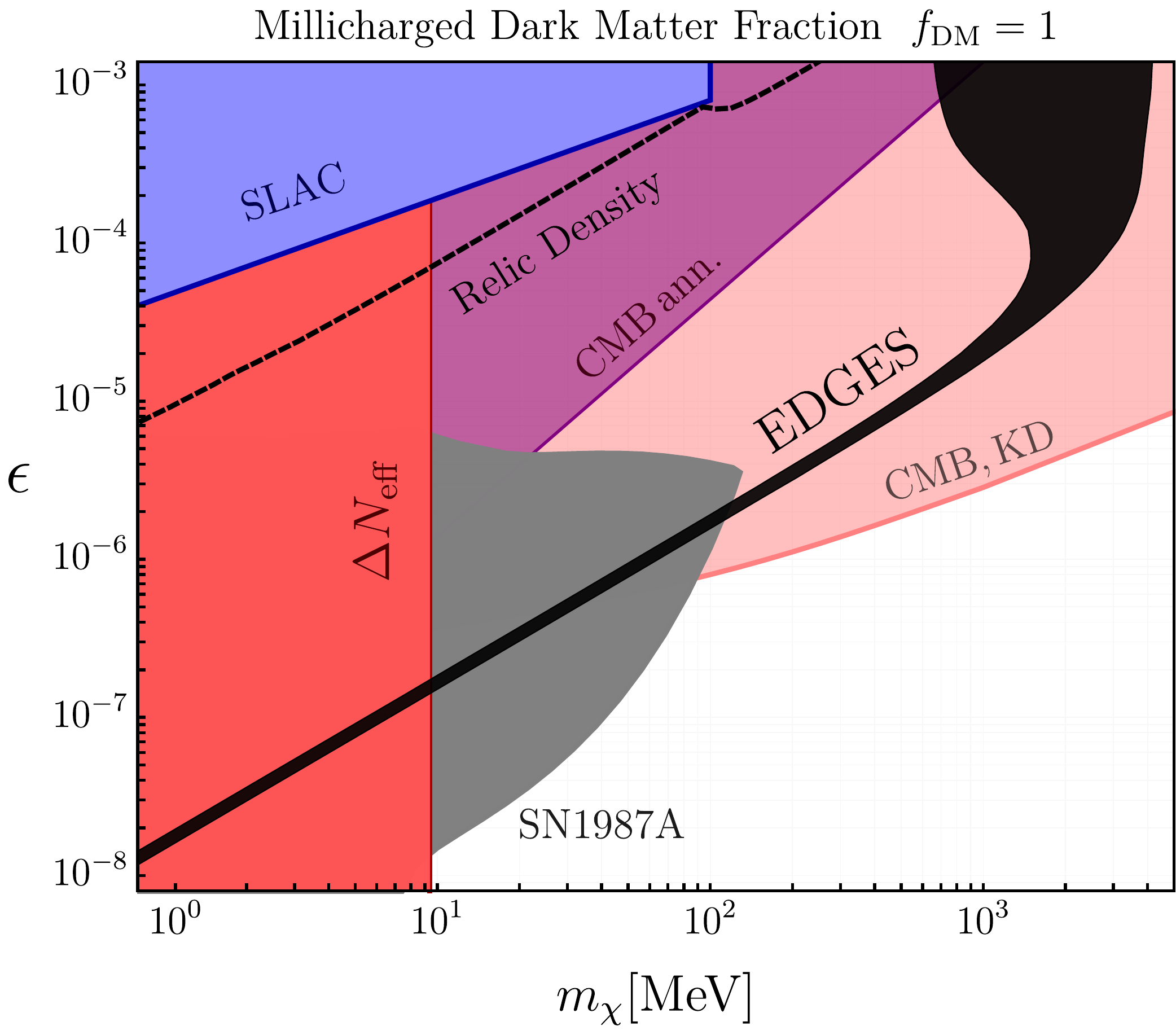}~~~~~~
\includegraphics[width=3.2in,angle=0]{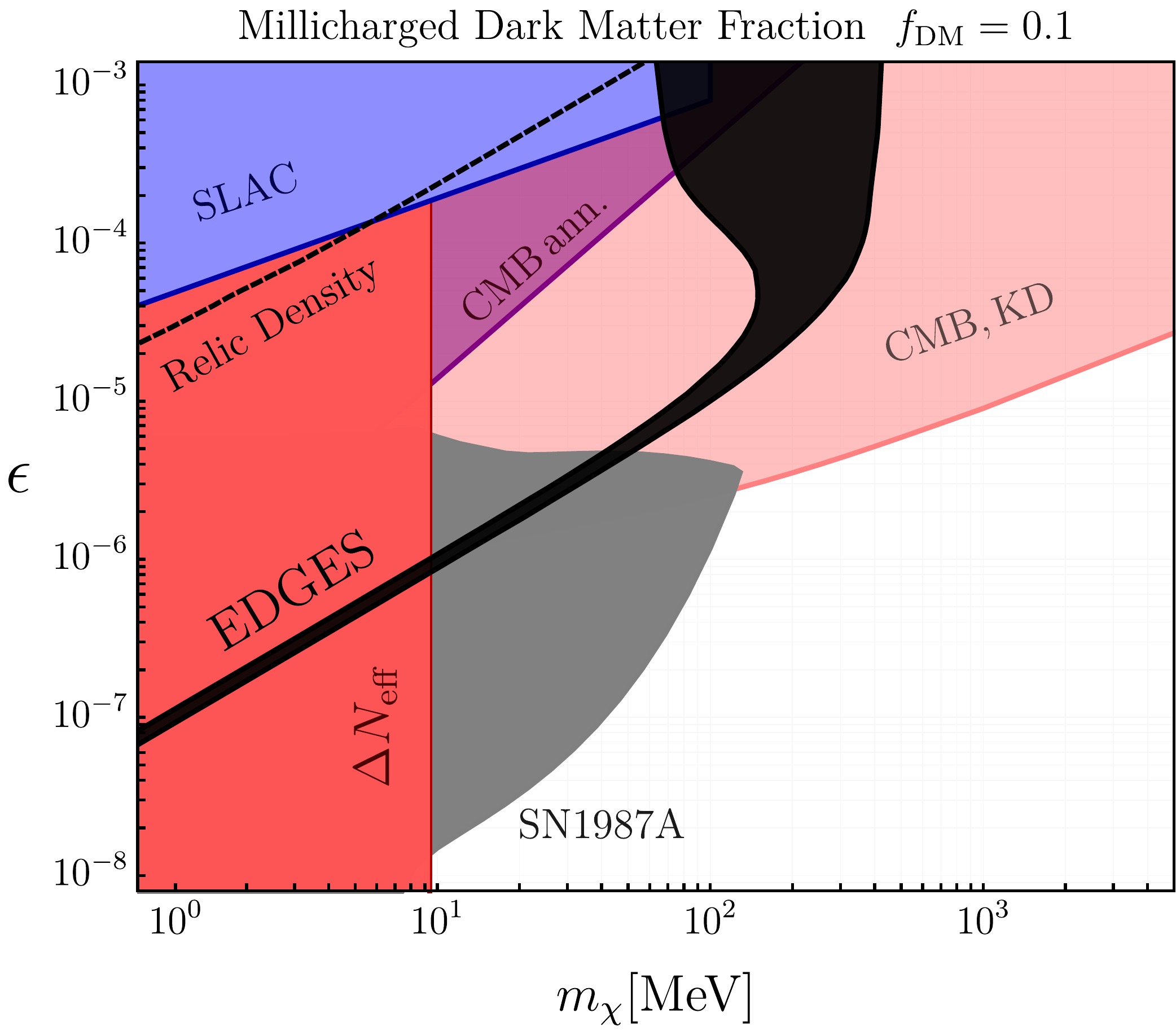} \\
\includegraphics[width=3.2in,angle=0]{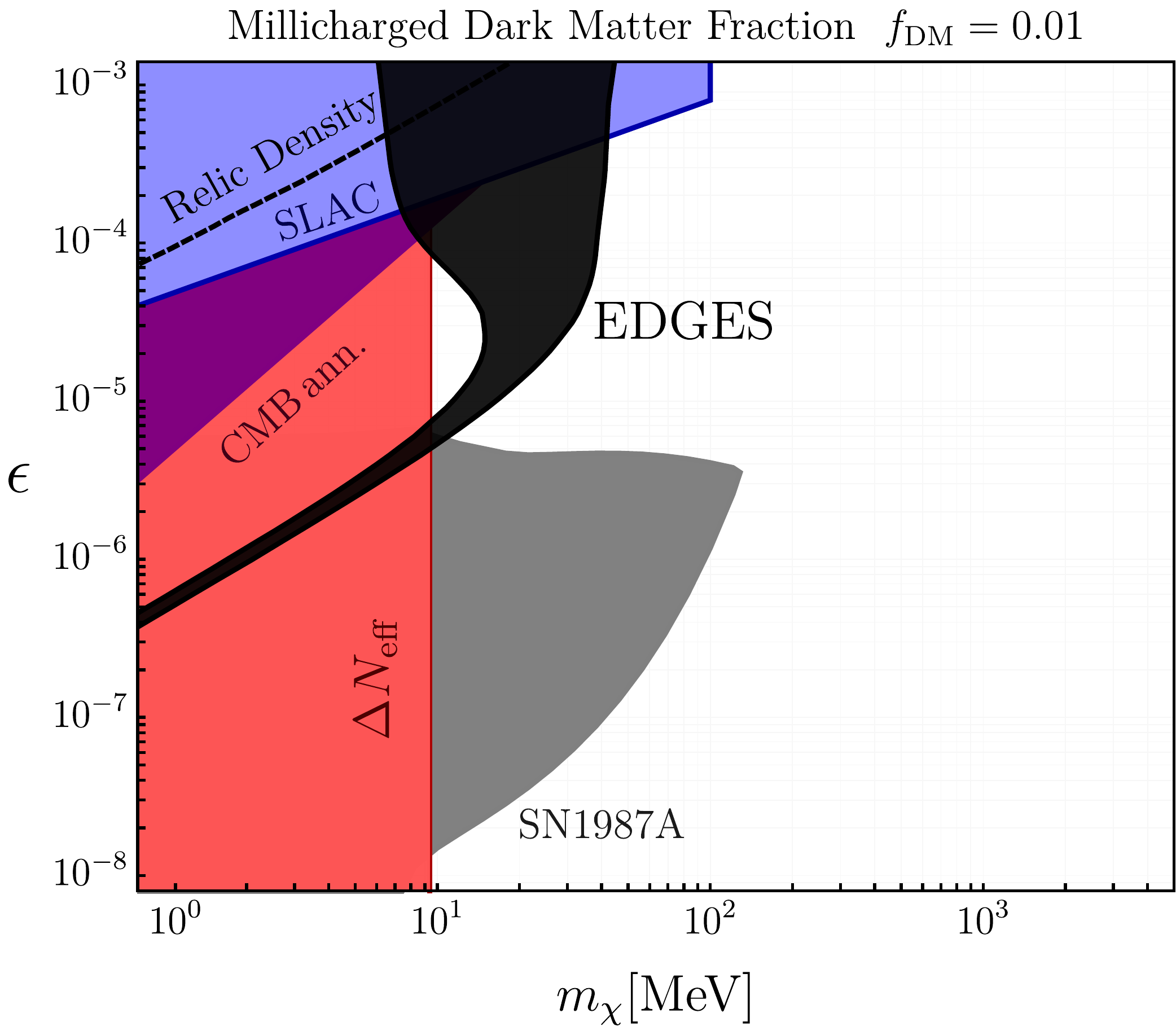}~~~~~~
\includegraphics[width=3.2in,angle=0]{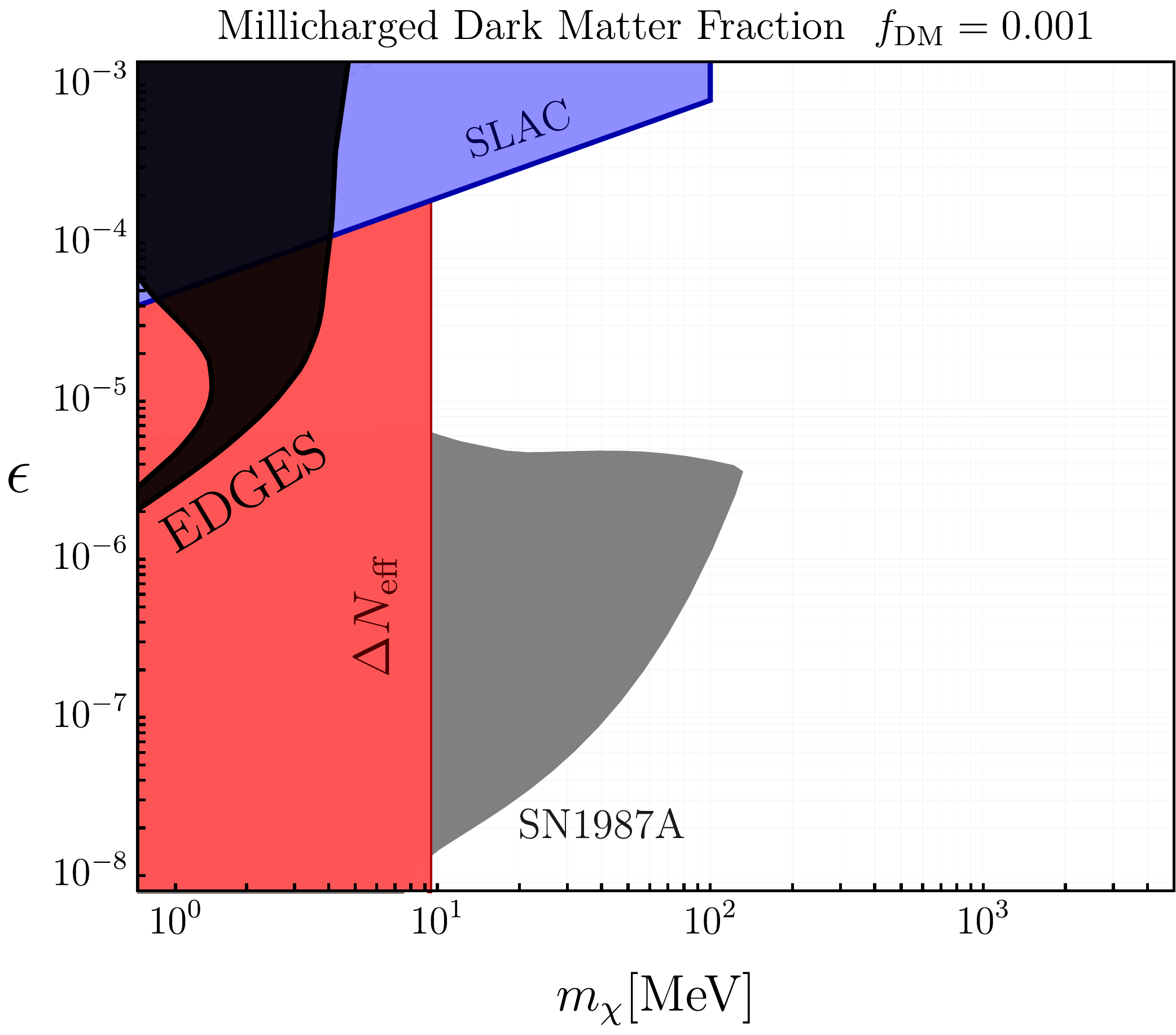} 
\caption{Constraints on Dirac fermion millicharged dark matter from Supernova 1987A (grey)~\cite{newsam}, the SLAC millicharge experiment (blue)~\cite{Prinz:1998ua}, the light element abundances produced during Big Bang Nucleosynthesis (red, labeled $\Delta N_{\rm eff}$)~\cite{Boehm:2013jpa}, and on the impact on the cosmic microwave background of dark matter scattering with baryons (pink, labeled CMB, KD)~\cite{McDermott:2010pa} and dark matter annihilations (purple, labeled CMB ann.)~\cite{Slatyer:2015jla}. These results are shown for four values of the fraction of the dark matter abundance that consists of millicharged particles, $f_{\rm DM}$. In each frame, the solid black regions represent the range of parameter values that could explain the amplitude of the observed 21-cm absorption feature as reported by the EDGES Collaboration~\cite{nature2}. The dashed black line denotes where the thermal relic abundance corresponds to quoted value of $f_{\rm DM}$, assuming only millicharge interactions. The fact that the solid black regions do not coincide with the dashed curves indicates that the dark matter must be depleted in the early universe by another kind of interaction. Although these results are shown for the specific case of dark matter in the form of a Dirac fermion, most of these constraints would change only very slightly if we were instead to consider a complex scalar. The exception to this are the constraints from dark matter annihilation during the epoch of recombination~\cite{Slatyer:2015jla}, which are much weaker in the complex scalar case, due to the $p$-wave suppression of the annihilation cross section.}
\label{constraints}
\end{figure*}

\section{Model-Independent Constraints On Millicharged Dark Matter}

Based on the results presented in Ref.~\cite{nature2}, it follows that millicharged dark matter particles could cool the gas at $z \sim 17$ to a level consistent with the EDGES measurement ($T_b \approx 4$ K) if the following condition is met~\cite{Munoz:2018pzp}:
\be
\epsilon \approx 1.7\times 10^{-4} \bigg(\frac{m_{\chi}}{300 \,{\rm MeV}}\bigg) \bigg(\frac{10^{-2}}{f_{\rm DM}}\bigg)^{3/4},
\label{condition}
\ee
where $\epsilon \equiv e_{\chi}/e$ is the electric charge of the dark matter, $m_{\chi}$ is the mass of the millicharged dark matter candidate, and $f_{\rm DM}$ is the fraction of the dark matter that consists of millicharged particles. This expression is valid for $m_{\chi} \lsim (20-40) \, {\rm MeV} \times (f_{\rm DM}/10^{-2})$, above which much larger values of $\epsilon$ are required. We also note that this expression applies to millicharged dark matter in either the form of a Dirac fermion or a complex scalar.

Millicharged dark matter is subject to a wide range of experimental and astrophysical constraints, some of which we summarize in Fig.~\ref{constraints}. Here, we show constraints derived from observation of Supernova 1987A (grey)~\cite{newsam}, from the SLAC millicharge experiment (blue)~\cite{Prinz:1998ua}, and from the light element abundances produced during Big Bang Nucleosynthesis (BBN), assuming entropy transfer to electrons and photons (red)~\cite{Boehm:2013jpa} (see also Refs.~\cite{Nollett:2014lwa,Steigman:2014uqa}). We note that the BBN constraints shift only at the tens of percent level if we had instead considered entropy transfers to neutrinos. We also show constraints from measurements of the CMB based on dark matter annihilation in the epoch of recombination (purple)~\cite{Slatyer:2015jla} (see also Ref.~\cite{Ade:2015xua}), and on dark matter scattering with baryons (pink)~\cite{McDermott:2010pa}. Note that we have modified the constraint from Ref.~\cite{McDermott:2010pa} by rescaling the limits on $\epsilon$ by a factor of $\sqrt{f_{\rm DM} \Omega_{\rm dm}/\Omega_b}$. We also note that Ref.~\cite{Xu:2018efh} has recently presented an even more stringent constraint on low-velocity dark matter-baryon scattering. Although these results are shown for the specific case of dark matter in the form of a Dirac fermion, most of these constraints would change only very slightly if we were instead to consider a complex scalar. The exception to this are those constraints from dark matter annihilation during recombination~\cite{Slatyer:2015jla}, which are much weaker in the case of a complex scalar, due to the $p$-wave suppression of the annihilation cross section.

Analytical constraints on dark matter scattering with baryons during recombination were presented in Ref.~\cite{McDermott:2010pa} for the case of $f_{\rm DM} = 1$. For $f_{\rm DM} \gtrsim 0.02$, these constraints can be straightforwardly rescaled. For smaller values of $f_{\rm DM}$, however, one cannot apply this bound, because the energy density of this component of the dark matter is smaller than the difference between the (95\% CL) upper limit on the baryon density from CMB~\cite{Ade:2015xua} and the (95\% CL) lower limit on the baryonic density based on BBN~\cite{Cyburt:2015mya}. In this case, the millicharged particles themselves could contribute to the apparent baryonic density as derived from the CMB, evading the CMB constraint even if tightly coupled to the baryon fluid. For this reason, the CMB kinetic decoupling bounds do not apply for $f_{\rm DM} \lesssim 0.02$.

The solid black regions in Fig.~\ref{constraints} represent the parameter space in which the reported amplitude of the 21-cm absorption feature can be explained. For $f_{\rm DM}=1$, we take this region as presented in Ref.~\cite{nature2}. For $f_{\rm DM} < 1$, we shift these regions in $\epsilon$ and $m_{\chi}$ as found in the numerical solutions presented in Ref.~\cite{Munoz:2018pzp} (see Eq.~\ref{condition}). 

In light of the constraints presented in Fig.~\ref{constraints}, we are forced to consider millicharged dark matter in a relatively narrow range of parameter space:  $m_{\chi} \sim 10-80$ MeV, $\epsilon \sim 10^{-6}-10^{-4}$, and $f_{\rm DM} \sim 0.003-0.02$. For the remainder of this paper, we will focus on this range of scenarios. We will not assume that the millicharge is accompanied by an extremely light dark photon \cite{Izaguirre:2015eya}.

We note that although a millicharged particle with a mass and coupling in this range could not have free-streamed out of Supernova 1987A, and thus was not deemed to be ruled out in Ref.~\cite{newsam}, such a particle would be in chemical equilibrium in the proto-neutron star. By equipartion, this would lower the temperature of the supernova core, with likely ramifications for the observed neutrino signal. It is possible that these considerations could be used to rule out this range of parameter space, although no high-resolution numerical studies have definitively addressed this issue within this class of scenarios.

\section{Depleting the Dark Matter Abundance}

A major challenge for millicharged dark matter models which can explain the reported amplitude of the observed 21-cm absorption feature is avoiding overproduction in the early universe. The annihilation cross section for a pair of millicharged particles is given by:
\be
\sigma v_{\chi \bar \chi \to  f\bar f}= \frac{\pi \alpha^2 \epsilon^2}{m^2_{\chi}} \, \kappa \, \bigg(1-\frac{m^2_f}{m^2_{\chi}}\bigg)^{1/2} \, \bigg(1+\frac{m^2_f}{2m^2_{\chi}}\bigg),
\label{annsigma}
\ee
where $\kappa= 1$ or $v^2/6$ (where $v$ is the relative velocity of the dark matter particles) for dark matter in the form of a Dirac fermion or a complex scalar, respectively. 
The dark matter will be maintained in chemical equilibrium with the Standard Model bath if its annihilation rate exceeds that associated with Hubble expansion. Performing this comparison at $T\sim m_\chi$ (when the rate is approximately maximal) yields the following condition for equilibrium:
\be
 n_\chi  \sigma v_{\chi \bar \chi \to  f\bar f} \sim  m_\chi^3 \frac{\pi \epsilon^2 \alpha^2 }{ m_\chi^2 } \, \kappa  > 1.66 \sqrt{g_*}\frac{m_\chi^2}{M_{\rm Pl}}, ~~~
\ee
where $g_{*}$ is the number of relativistic degrees of freedom in the thermal bath, and $M_{\rm Pl}$ is the Planck mass. This in turn implies that if
\be
\epsilon \gtrsim \frac{g_*^{1/4}}{\alpha} \sqrt{\frac{m_\chi}{m_{\rm Pl}}} \, \frac{1}{\kappa^{1/2}}\sim 10^{-8} \left( \frac{m_\chi}{ \rm 10~ MeV} \right)^{1/2} \! \frac{1}{\kappa^{1/2}},~~
\ee
then $n_{\chi}$ will reach its equilibrium value in the early universe, and thus must be depleted through efficient annihilations in order to avoid exceeding the desired density of millicharged particles, $f_{\rm DM}  \Omega_{\rm CDM}$. In particular, equilibrium is reached over the entirety of the parameter space for which dark matter can explain the amplitude of the observed 21-cm absorption feature. 

Once equilibrium is reached, the thermal relic abundance is determined by the dark matter's annihilation cross section. For the cross section given in Eq.~\ref{annsigma}, this leads to a thermal abundance that makes up the following fraction of the dark matter density: $f_{\rm DM} \approx 0.04 \times (10^{-3}/\epsilon)^2 \times (m_{\chi}/30\, {\rm MeV})^2 \times \kappa^{-1}$~\cite{Steigman:2012nb}. In Fig.~\ref{constraints}, the dashed black lines indicate the regions of parameter space in which the thermal relic abundance corresponds to quoted value of $f_{\rm DM}$, assuming only millicharge interactions. The fact that the solid regions do not coincide with the dashed curves in the allowed regions of millicharge parameter space indicates that the dark matter must be depleted by an interaction other than annihilations through photon exchange.

To deplete the thermal abundance of millicharged dark matter to an acceptable level, we consider the possibilities of either annihilations directly to Standard Model particles or to particles within a hidden sector. In either case, we require such annihilations to take place through $p$-wave processes, as measurements of the CMB rule out dark matter candidates lighter than $\sim$10 GeV if they freeze-out primarily through $s$-wave annihilations (with the exception of annihilation to neutrinos)~\cite{Slatyer:2015jla}.

\subsection{Annihilation to Standard Model Fermions}

\begin{figure}
\hspace{-0.2in}
\includegraphics[width=3.4in,angle=0]{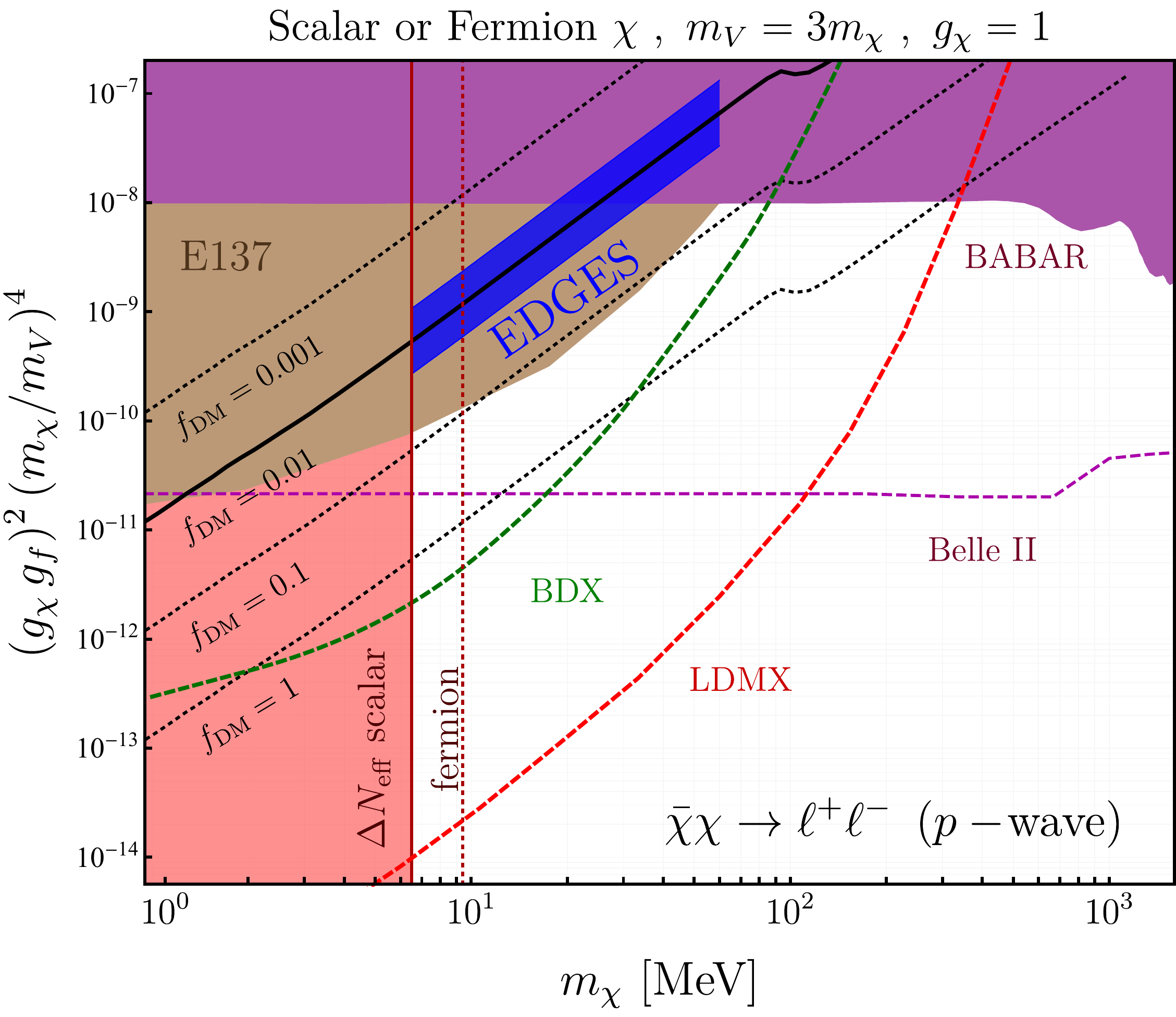}
\caption{Constraints on a scenario in which millicharged dark matter, $\chi$, annihilates to Standard Model leptons through the exchange of a vector mediator, $V$, that couples universally to all three generations of charged leptons. The blue band denotes the approximate region of parameter space which can explain the amplitude of the observed 21-cm absorption feature as reported by the EDGES Collaboration. This case is ruled out by constraints from the BABAR~\cite{Lees:2017lec,Izaguirre:2013uxa,Essig:2013vha} and E137~\cite{Bjorken:1988as,Izaguirre:2013uxa} experiments, along with BBN, over the entire range of parameter space that is capable of generating the reported 21-cm signal. Also shown as dashed lines are the projected constraints of the  Belle II~\cite{Izaguirre:2014bca}, BDX~\cite{Izaguirre:2013uxa,Battaglieri:2016ggd} and LDMX~\cite{Izaguirre:2014bca,Battaglieri:2017aum} experiments.  Here we have adopted $m_{V}=3m_{\chi}$ and $g_{\chi}=1$. We note that for other choices of this coupling and mass ratio, the constraints are generically more restrictive than those presented here.}
\label{direct}
\end{figure}

\begin{figure*}
\includegraphics[width=3.4in,angle=0]{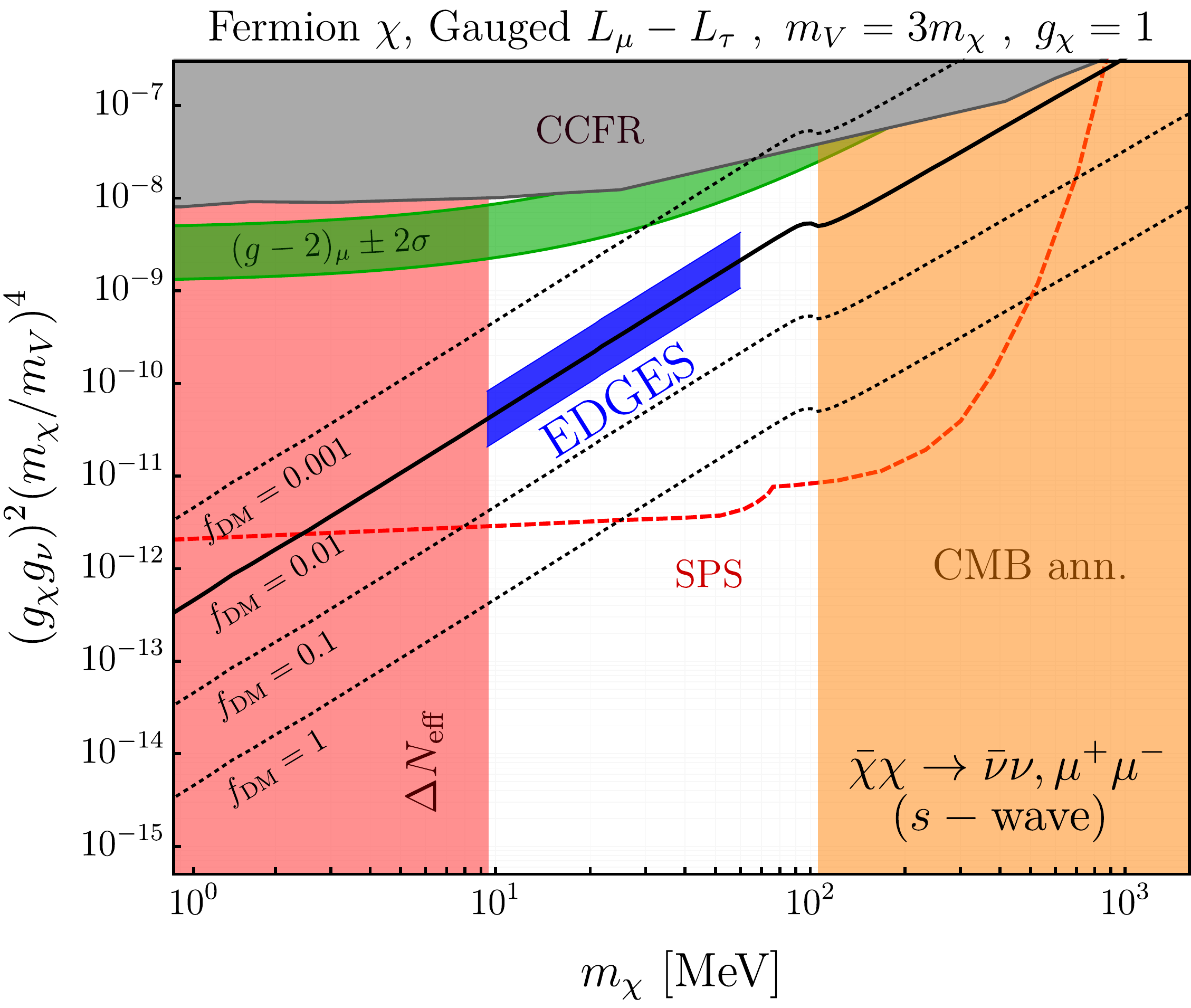}~~~~~
\includegraphics[width=3.4in,angle=0]{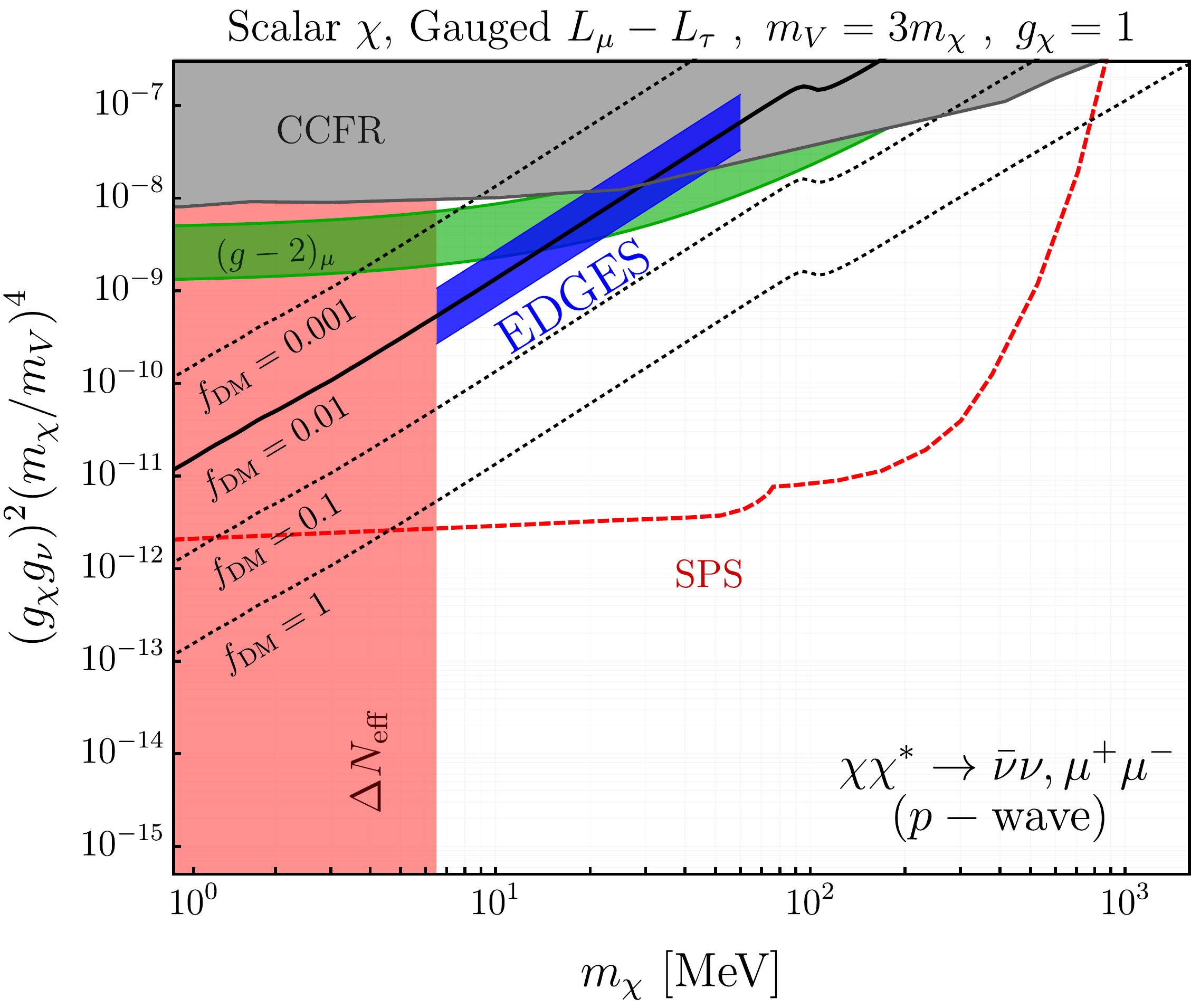}
\caption{Constraints on a scenario in which millicharged dark matter, $\chi$, annihilates to Standard Model leptons through the exchange of a vector mediator, $V$, which couples to muon minus tau number, associated with the gauge group $U(1)_{L_{\mu}-L_{\tau}}$. The blue band denotes the approximate region of parameter space which can explain the amplitude of the observed 21-cm absorption feature as reported by the EDGES Collaboration. Because this mediator does not directly couple to electrons, the experimental constraints are considerably less restrictive in this case, potentially allowing the dark matter to be sufficiently depleted in the early universe. Future measurements by SPS~\cite{Gninenko:2014pea} are expected to be sensitive to this scenario. Also shown are the regions that are capable of explaining the observed value of the muon's magnetic moment, $(g-2)_{\mu}$~\cite{Bennett:2006fi,Hagiwara:2011af,Davier:2010nc,Jegerlehner:2009ry}, as well as those that are ruled out by the CCFR experiment~\cite{Altmannshofer:2014pba,Mishra:1991bv}. In each frame, we have adopted $m_{V}=3m_{\chi}$ and $g_{\chi}=1$.}
\label{direct2}
\end{figure*}

For the case of annihilation directly to Standard Model states, we consider two options for $p$-wave processes: dark matter in the form of a scalar or a fermion which annihilates through a new vector, $V$ (fermionic annihilation through a scalar mediator also leads to $p$-wave amplitudes, but is more constrained). In both of these cases, the vector must be heavier than the dark matter itself (in order for annihilations to Standard Model fermions to dominate over those to $VV$). First, we consider the following interactions which lead to $p$-wave suppressed annihilations:
\begin{subequations}
\begin{eqnarray}
\mathcal{L}_f  &\supset& V_\mu ( g_\chi \bar \chi \gamma^\mu \gamma^5 \chi  + g_f \bar f \gamma^\mu f), \label{lagp-f} \\
\mathcal{L}_s  &\supset& V_\mu (i g_\chi \chi^* \partial_\mu \chi +g_f  \bar f \gamma^\mu f  + {\rm h.c.}),  \label{lagp-s}
\end{eqnarray}
\end{subequations}
where the dark matter candidate, $\chi$, is a scalar or a fermion, respectively. Although for a Dirac fermion it is also possible to have $p$-wave $\chi \bar \chi \to f \bar f$ annihilation through scalar $s$-channel exchange, this possibility is strongly excluded in simple models \cite{Krnjaic:2015mbs}. For the model of Eq.~\ref{lagp-f} or \ref{lagp-s}, we can write the annihilation cross section as follows:
\be
 \sigma v = \frac{g^2_f g^2_{\chi} m^2_{\chi} v^2}{6 \pi (4m^2_{\chi}-m^2_V)^2}, 
\ee
where we have taken the $m_\chi \gg m_f$ limit.

In Fig.~\ref{direct}, we plot constraints on models defined as in Eqs.~\ref{lagp-f} and \ref{lagp-s} in which  the millicharged dark matter annihilates through a vector mediator that couples universally to all three species of charged leptons (for details, see Ref.~\cite{Izaguirre:2015yja}). In this case, constraints from the BABAR~\cite{Lees:2017lec,Izaguirre:2013uxa,Essig:2013vha} and E137~\cite{Bjorken:1988as,Izaguirre:2013uxa} experiments, along with BBN, exclude the entire range of parameter space that is capable of generating the reported 21-cm signal ($m_{\chi} \sim 10-80$ MeV and $f_{\rm DM} \sim 0.003-0.02$).

We also consider scenarios in which dark matter can annihilate predominantly to neutrinos (as opposed to $e^+e^-$), with either $s$- or $p$-wave annihilation:
\begin{subequations}
\begin{eqnarray}
\mathcal{L}_f &\supset&  V_\mu ( g_\chi \bar \chi \gamma^\mu \chi  + g_{\nu} \bar \nu_L \gamma^\mu \nu_L ), \label{lags-f}
\\
\mathcal{L}_s  &\supset& V_\mu ( i g_\chi \chi^* \partial_\mu \chi  + g_{\nu} \bar \nu_L \gamma^\mu \nu_L  + {\rm h.c.} ). \label{lags-s}
\end{eqnarray}
\end{subequations}
These models have the annihilation cross section,
\be
 \sigma v = \frac{g^2_{\nu} g^2_{\chi} m^2_{\chi} \kappa}{2 \pi (4m^2_{\chi}-m^2_V)^2},
\ee
to a given neutrino flavor where $\kappa=1$ or $v^2/6$ for a fermion or scalar, respectively. A concrete example of this would be a model in which the vector is associated with the anomaly-free gauge group $U(1)_{L_{\mu}-L_{\tau}}$ and thus couples to muons, taus, and their respective neutrino species~\cite{Baek:2001kca,Pospelov:2008zw}. We show in Fig.~\ref{direct2} the parameter space of such a model for either a fermionic dark matter candidate with interactions as defined in Eq.~\ref{lags-f} or a scalar dark matter candidate with interactions as defined in Eq.~\ref{lags-s}. (If we had included a $\gamma_5$ in  Eq.~\ref{lags-f} to make the annihilation $p$-wave suppressed, the results would be almost indistinguishable from the scalar case.) Because this mediator does not directly couple to electrons, the experimental constraints are considerably less restrictive in this case, potentially allowing the dark matter to be sufficiently depleted in the early universe. We note that in the scalar dark matter case, this model predicts a contribution to the magnetic moment of the muon, $(g-2)_{\mu}$, that is capable of explaining the measured anomaly~\cite{Bennett:2006fi,Hagiwara:2011af,Davier:2010nc,Jegerlehner:2009ry}. The same is true in the fermionic case, but for a somewhat lower value of $g_{\chi}$. It is anticipated that future measurements by SPS~\cite{Gninenko:2014pea} will be sensitive to this scenario~\cite{gordan}.

\subsection{Annihilation to Hidden Sector Particles}

Next we turn our attention to scenarios in which the dark matter annihilates to unstable particles that reside within a hidden sector. In light of the $p$-wave requirement, we focus on annihilations to a pair of real scalars, $\phi$, which we take to be lighter than the dark matter itself. We motivate this choice by the fact that a real scalar leads to the smallest contribution to the energy density of dark radiation (\ie~the number of effective neutrino species, $N_{\rm eff}$). The dark matter annihilation cross section to a pair of these dark scalars is given by:
\be
 \sigma v _{\chi \bar \chi \to \phi \phi} = \frac{3 y^4 v^2}{128 \pi m_\chi^2}, 
\ee
where $y$ is the $\phi \bar \chi \chi$ yukawa coupling and $v$ is the relative velocity of the annihilating particles. This cross section leads to a relic abundance equal to the following fraction of the measured dark matter abundance:
\be
f_{\rm DM} \approx 0.008 \, \bigg(\frac{0.03}{y}\bigg)^4 \, \bigg(\frac{m_{\chi}}{30 \, {\rm MeV}}\bigg)^2 \left( \frac{0.1}{v^2} \right),
\ee
where we assume that a thermal relic Dirac fermion attains the observed abundance with $ \sigma v \simeq 6\times 10^{-26} {\rm cm^3/s}$.

The dark matter models under consideration here are strongly constrained by observations of the CMB and the successful predictions of BBN (see, for example, Refs.~\cite{Vogel:2013raa,Brust:2013xpv}). If the $\phi$ population is relativistic during BBN, it will contribute $\Delta N_{\rm eff} = 4/7$, in excess of the range of values allowed by measurements of the light element abundances~\cite{Cyburt:2015mya}. On the other hand, if the $\phi$ population is non-relativistic and decays after BBN, this will ruin the concordance of the baryon-to-photon ratio as measured during BBN and recombination.

In order to evade this constraint, the $\phi$ abundance must be transferred prior to BBN into Standard Model particles, which then reach equilibrium with both photons and neutrinos, thereby preserving the Standard Model prediction $\neff = 3.046$~\cite{Mangano:2001iu}. This could be facilitated, for example, through the rapid decay of the $\phi$. Unless we include additional particle content into our model, the $\phi$ population will decay predominantly to a pair of photons through a $\chi$ loop, with a width that is given by:
\be
\Gamma_{\phi \rightarrow \gamma \gamma} =  \frac{y^2 \epsilon^4 \alpha^2 m^3_{\phi}}{256\pi^3 m^2_{\chi}} \, \left|A_{1/2}(m_{\phi},m_{\chi}) \right|^2,
\ee
where
\be
A_{1/2} \equiv \frac{8 m^2_{\chi}}{m^2_{\phi}} \, \bigg[1+\bigg(1-\frac{4 m^2_{\chi}}{m^2_{\phi}}\bigg) \arcsin^2 \left(\frac{m_\phi}{2m_\chi} \right) \bigg]. \,\,\,\,\,\,
\ee
We find that there is no combination of parameters that can account for the observed amplitude of the 21-cm absorption feature for which this decay will take place prior to BBN ($\tau_{\phi} \lsim 1$ s). Thus to avoid the stringent constraints from BBN, we must introduce an additional mechanism to more rapidly deplete the $\phi$ abundance, such as through the decay to Standard Model fermions through mixing with the Higgs. Alternatively, we could also consider interactions which deplete the $\phi$ abundance through 3-to-2 processes~\cite{Carlson:1992fn,Hochberg:2014kqa,Pappadopulo:2016pkp}.

We have not been entirely exhaustive in considering the possibilities for depleting the density of millicharged dark matter in the early universe. Other model building options include annihilation to ``forbidden''~\cite{DAgnolo:2015ujb} or ``impeded''~\cite{Kopp:2016yji} final states, which can evade CMB constraints by suppressing annihilation at low-velocities; processes in which the dark matter freezes out through 4-to-2 processes~\cite{Herms:2018ajr}; and scenarios featuring additional reheating after dark matter freeze-out but prior to BBN~\cite{Berlin:2016gtr,Randall:2015xza,Davoudiasl:2015vba,Gelmini:2006pq,Kane:2015jia}.

\section{Summary and Conclusions}

The recently reported detection by the EDGES Collaboration of a feature at 78 MHz in the sky-averaged radio spectrum marks a potentially momentous occasion for astrophysics. The possibility that such a signal is not compatible with standard astrophysics is even more exciting, potentially pointing to a strong coupling between the baryons and dark matter at a redshift of $z \sim 17$. This striking possibility merits substantial investigation and scrutiny.

In this paper, we have pointed out several very stringent constraints on dark matter candidates potentially capable of producing the reported 21-cm feature. Since a mediator that communicates a new long-range force with couplings of the required size is ruled out by both laboratory and cosmological considerations, we are forced to consider models in which the dark matter itself possesses a small electric charge. Such models are highly constrained, however, by measurements of the cosmic microwave background, light element abundances, and Supernova 1987A. In particular, if such a particle constitutes more than $2\%$ of the dark matter, it would damp baryon density perturbations at the time of recombination to an unacceptable degree. After applying these constraints, we find that the only range of models that could potentially explain the reported 21-cm signal are those in which a small fraction $\sim0.3-2\%$ of the dark matter consists of particles with a mass of $\sim10-80$ MeV and which couple to the photon through a small electric charge of $\epsilon \sim 10^{-6}-10^{-4}$.

Furthermore, throughout this range of models, the dark matter is predicted to have reached thermal equilibrium with the Standard Model in the early universe, thus requiring that the model be supplemented with an additional mechanism to deplete the dark matter abundance to an acceptable level. We consider scenarios in which the dark matter annihilates through a new vector to neutrinos, or to particles within a hidden sector which are themselves depleted through rapid decays or 3-to-2 processes. Specific possibilities include annihilations to neutrinos through a $U(1)_{L_{\mu}-L_{\tau}}$ gauge boson, which is expected to be within the reach of future measurements by the SPS experiment.

\bigskip
\begin{acknowledgments}  
We would like to thank Kimberly Boddy, Cora Dvorkin, Rouven Essig, Vera Gluscevic, Wayne Hu and Marc Kamionkowski for helpful communication. Some of the results presented in this work utilized PackageX~\cite{Patel:2015tea}. AB is supported by the U.S. Department of Energy under Contract No. DEAC02-76SF00515. This manuscript has been authored by Fermi Research Alliance, LLC under Contract No. DE-AC02-07CH11359 with the U.S. Department of Energy, Office of Science, Office of High Energy Physics. The United States Government retains and the publisher, by accepting the article for publication, acknowledges that the United States Government retains a non-exclusive, paid-up, irrevocable, world-wide license to publish or reproduce the published form of this manuscript, or allow others to do so, for United States Government purposes.
\end{acknowledgments}

\bibliography{21cm}

\end{document}